\documentstyle[12pt]{article}
\begin{document}
\baselineskip=7mm

\noindent
{\bf SAGA-HE-120-97}

\noindent
{\bf March 1997}

\bigskip

\bigskip

\centerline{\bf Vacuum Effects and Compressional Properties of Nuclear Matter}
\centerline{\bf in Cutoff Field Theory}

\bigskip

\centerline{By}

\noindent{\bf Hiroaki Kouno$^*$, Katsuaki Sakamoto, Yoshitaka Iwasaki, }

\noindent{\bf Nobuo Noda, Tomohiro Mitsumori, Kazuharu Koide,  Akira Hasegawa}

\centerline{Department of Physics, Saga University, Saga 840, Japan}

\centerline{and}

\centerline{\bf Masahiro Nakano}

\centerline{University of Occupational and Environmental Health, Kitakyushu 807, Japan}

\bigskip

\noindent
* e-mail address: kounoh@cc.saga-u.ac.jp

\centerline{\bf PACS numbers: 21.65.+f}

\bigskip

\bigskip

\centerline{\bf Abstract}

\bigskip

\noindent
Including the vacuum effects, the compressional properties of nuclear matter are studied in the cutoff field theory. 
Under the Hartree approximation, the low-energy effective Lagrangian is derived in the framework of the renormalization group methods. 
The coefficients are determined in a way where the physical results hardly depend on the value of the cutoff which is conveniently introduced into the theory. 
It is shown that, to reproduce the empirical data of the nucleus incompressibility, the compressibility of the nuclear matter is favorable to be 250$\sim$350MeV. 

\vfill\eject



\centerline{\bf 1. Introduction}

\bigskip

In recent two decades, nuclear matter has been studied in the framework of quantum hadrodynamics (QHD). 
The meson mean-field theory for nuclear matter [1] has made successful results to account for the saturation properties at the normal nuclear density. 
Following to those successes, many studies and modifications are done in the relativistic nuclear models. 
One of those modifications is inclusion of vacuum fluctuation effects, which cause divergences in physical quantities as they are naively calculated. 
Chin [2] estimated the vacuum fluctuation effects in the Hartree approximation by using the renormalization procedures, and found that the vacuum fluctuation effects make the incompressibility of nuclear matter smaller and closer to the empirical value than in the original Walecka model. 
However, it becomes more difficult to do the renormalization as the model becomes more complicated, since the renormalization procedures need analytical studies to some extent. 
Numerical studies using the cutoff field theory may be useful. 

Furthermore, the relation between QHD and the underlying fundamental theory, i.e., QCD, is an open question. 
One may wonder whether QHD is valid in very high-energy scale or not. 
If QHD is valid only under some energy scale, it is natural to introduce a cutoff or a form factor into the theory. 
One may introduce the cutoff [3] or the form factor [4] to avoid the instability of the meson propagators  in the random phase approximation (RPA) [5]. 
Cohen [6] introduced the four dimensional cutoff into the relativistic Hartree calculation and found that the vacuum energy contribution may be somewhat different from the one in the renormalization procedures, if the cutoff is not so large. 

In the recent paper [7], we have studied the nuclear matter properties in details using the cutoff field theory of Cohen and show that the properties of nuclear matter may be somewhat different from those described in the ordinary renormalization procedures if the cutoff is not so large. 
In sec. 3 of the same paper, the cutoff field theory of the nuclear matter is reformulated in the framework of the renormalization group methods [8][9][10] and the fifth and sixth order terms of $\sigma$-meson self-interactions in the effective Lagrangian are shown to be important in the cutoff field theory. 
We have also studied the meson masses at finite baryon density [11] and the vertex corrections [12] in the framework of the cutoff field theory and renormalization group methods. 
In this paper, we studies the compressional properties of the nuclear matter in detail by using the method presented in sec. 3 of [7]. 

This paper is organized as follows. 
In section 2, we review our method to study the vacuum effects in the cutoff field theory. 
In section 3, we show the numerical analyses of the compressional properties of the nuclear matter. 
The section 4 is devoted to the summary.


\bigskip

\bigskip

\centerline{\bf 2. Cutoff Field Theory for Nuclear Matter}

\bigskip

At first, 
we start with the following renormalizable Lagrangian of $\sigma$-$\omega$ model together with a regulator that truncates the theory's state space at some large $\Lambda$. 
$$
L={\bar{\psi}}(i\gamma_\mu\partial^\mu-M+g_s\phi-g_v\gamma_\mu V^\mu )\psi
+{{1}\over{2}}\partial_\mu\phi\partial^{\mu}\phi-{{1}\over{2}}m_s^2\phi^2
-U(\phi )$$
$$
-{{1}\over{4}}F_{\mu\nu}F^{\mu\nu}+{{1}\over{2}}m_v^2V_\mu V^\mu
;$$
$$ U(\phi )=\sum_{n=0}^{4}C_n(g_s\phi )^n, \eqno{(1)} $$
where $\psi$, $\phi$, $V_\mu$, $M$, $m_s$, $m_v$, $g_s$, and $g_v$ are nucleon field, $\sigma$-meson field, $\omega$-meson field, nucleon mass, $\sigma$-meson mass, $\omega$-meson mass, $\sigma$-nucleon coupling, and $\omega$-nucleon coupling, respectively. 
The $C_n$ are constant parameters which are adjusted to reproduce the physical conditions as explained below. 
The Lagrangian (1) is valid only in the region of the energy scale which is smaller than $\Lambda$. 

In the relativistic Hartree approximation with the cutoff $\Lambda$, the one-loop contribution to the $\sigma$ effective potential is given by [6][7]
$$ U_{1-loop}(M^*,\Lambda )=-\int {{d^4k_E}\over{(2\pi)^4}}{{1}\over{2}}{\rm Tr}\big( \log{\big( {{k_E^2+M^{*2}}\over{\mu^2}} \big) }\big) \Theta (\Lambda^2-k_E^2), \eqno{(2)} $$
where $M^*=M-\Phi=M-g_s<\phi >$, $\mu$ is an arbitrary scale parameter with dimensions of mass, $\Theta$ is the step function, and the subscript $E$ denotes that the momentum with it is written in the Euclidian notation. 
(We remark that $\mu$-dependence appears only in the density-independent term in the effective potential (2). 
So the $\mu$-dependence of the energy density $\varepsilon$ also appears only in the density-independent term of $\varepsilon$ and disappears when we choose the coefficient $C_0$ to ensure $\varepsilon =0$ at zero density, as is described below. )
We choose parameters $C_n$ to reproduce the following conditions as in the ordinary renormalization procedures [2][13]. 
$$ {{d^n}\over{d\Phi ^n}}[U(\Phi )+U_{1-loop}(\Phi )]\vert_{\Phi=0}=0.~~~~(n=0,1,2,3,4) \eqno{(3)} $$
Note that the different conditions give the different physical results as is pointed out by Heide and Rudaz [14]. 

The conditions give 
$$  C_n=-{1\over{n!}}U_{1-loop}^{(n)}(0)=-{1\over{n!}}{{d^n}\over{d\Phi^n}}U(\Phi)_{1-loop}\vert_{\Phi =0},~~~~~(n=0,1,2,3,4). \eqno{(4)} $$ 
The condition (4) with $n=0$ ensures that energy density $\varepsilon$ of the system becomes zero at zero baryon density ($\rho =0$ ) and remove the $\mu$-dependence of $\varepsilon$. 
The condition (4) with $n=1$ ensures that the scalar density $\rho_s$ of the nucleons becomes zero at $\rho=0$ and $M^*=M$ at $\rho=0$. 
The condition (4) with $n=2$ ensures that the physical mass of $\sigma$-meson is $m_s$. 
The conditions (4) with $n=3$ and $n=4$ mean that the effective cubic and quartic self-couplings of $\sigma$-meson vanish. 

After the $C_n$ are determined, the scalar part of the nucleon self-energy is calculated by the equation of motion (EOM) for $\sigma$-meson, i.e., 
$$ {{\partial (L-U_{1-loop})}\over{\partial \Phi}}=0.   \eqno{(5)} $$
The vector part of the nucleon self-energy is calculated by the equation of motion of $\omega$-meson as usual [1][2][13]. 
If we consider the limit $\Lambda\rightarrow \infty$, the cutoff model is equivalent to the ordinary model in the renormalization procedures [2][13]. 

In the Hartree approximation, the total energy density of the system is given by the following equation. 
$$   \varepsilon(k_F,M^*,\Lambda )
={{g }\over{(2\pi )^3}}\int^{k_F}_0dk^3\sqrt{{\bf k}^2+M^{*2}}
+{{C_v^2}\over{2M^2 }}\rho^2
+{{M^2}\over{2C_s^2}}\Phi^2
$$
$$+U(\Phi )+U_{1-loop}(M^* ,\Lambda ),     \eqno{(6)} $$
where $C_s=g_sM/m_s$, $C_v=g_vM/m_v$, $g$ is the degeneracy factor ($g=4$ in the symmetric nuclear matter ) and $k_F$ is the Fermi momentum. 
Note that eq. (5) is equivalent to 
$$   {{\partial \varepsilon (k_F,M-\Phi ,\Lambda )}\over{\partial \Phi}}=-{{\partial \varepsilon (k_F,M^*,\Lambda)}\over{\partial M^*}}=0.  \eqno{(7)} $$

In the sec. 2 of ref. [7], using the eqs. (6) and (7) with the coefficients (4) and the given value of the cutoff $\Lambda$, we have done self-consistent calculation in the Hartree approximation. 
However, here we use the method which is presented in the sec. 3 of ref. [7] and review the derivation of the low-energy effective Lagrangian in the framework of the renormalization group method [8][9][10]. 

If we do not know the value of $\Lambda$, the cutoff which we introduce into the theory (below we call this cutoff $\Lambda^\prime$ ) may be different from the true cutoff $\Lambda$ which is the limiting energy scale of the theory. 
Suppose $\Lambda^\prime$ is smaller than $\Lambda$. 
(Of course, $\Lambda^\prime$ should be larger than the energy scale of the physics in which we are  interested. ) 

In the renormalization group methods [8][9][10], we require that the physical quantities do not depend on $\Lambda^\prime$, although the physical quantities depend on $\Lambda$. 

To achieve this, we estimate the contributions which is needlessly discarded by introducing the cutoff $\Lambda^\prime$ which is smaller than $\Lambda$. 
In the case of the vacuum energy potential, it is given by
$$ \Delta U =-\Delta L \equiv U_{1-loop}(M^*,\Lambda^2\geq k^2_E \geq \Lambda^{\prime 2})$$
$$=U_{1-loop}(M^*,\Lambda )-U_{1-loop}(M^*,\Lambda^\prime)
$$
$$
=-\int {{d^4k_E}\over{(2\pi)^4}}{{1}\over{2}}{\rm Tr}[ \log{\big( {{k_E^2+M^{*2}}\over{\mu^2}} \big) }] [\Theta (\Lambda^2-k_E^2)-\Theta (\Lambda^{\prime 2}-k_E^2)]
$$
$$
=-{{g}\over{8\pi^2}}\int_{\Lambda^\prime}^{\Lambda}dk_Ek_E^3\log{\big( {{k_E^2+M^{*2}}\over{\mu^2}}\big)}
=-{{g\Lambda^{\prime4}}\over{8\pi^2}}\int_1^{\Lambda/\Lambda^\prime}dxx^3\log{\big( {{x^2+y^2}\over{(\mu/\Lambda^\prime )^2}}\big)}, \eqno{(8)} $$
where $x=k_E/\Lambda^\prime$ and  $y=M^*/\Lambda^\prime$. 
If $M^*/\Lambda^\prime <1$, we could expand the last line of eq. (8) around $y^2=0$. 
$$ \Delta U=\sum_{m=0}^{\infty}{{1}\over{m!}}{{\partial^m( \Delta U)}\over{\partial (y^{2})^m}}\vert_{y^2=0}(y^2)^{m}
=\sum_{m=0}^{\infty}{{1}\over{m!}}{{\partial^m (\Delta U)}\over{\partial (y^2)^{m}}}\vert_{y^2=0}\big( {{M-\Phi}\over{\Lambda^\prime}}\big) ^{2m}. \eqno{(9)} $$
Therefore, if we use the new cutoff $\Lambda^{\prime}$, we must add $\Delta L$ to the Lagrangian as the effective potential, to keep that physical qunatities do not depend on $\Lambda^\prime$. 
$L+\Delta L$ is the low energy effective Lagrangian under the energy scale $\Lambda^\prime$ in the Hartree approximation. 
With this effective Lagrangian, the energy density is obtained by
$$ \varepsilon (k_F, M^*, \Lambda^\prime )-\Delta L= \varepsilon (k_F, M^*, \Lambda^\prime )+\Delta U, \eqno{(10)} $$
which is equivalent to $\varepsilon (k_F,M^*,\Lambda )$. 
It should be remarked that, since the terms $\Phi^n$ with $n\leq 4$ can be absorbed in $U(\Phi )$ which is phenomenologically determined, only the terms with $n\geq 5$ in eq. (9) are essentially new. 

Since the coefficients of the expansion (9) are order $(\Lambda^\prime )^4$, the $m$-th term of (9) has the $(1/\Lambda^\prime )^{2m-4}$ order contribution if we treat $\Lambda$ as the same order as the $\Lambda^\prime$. 
In actual calculations, we truncate the Taylor expansion (9) at some finite maximum $m$. 
In that case, the calculated results include $O((1/\Lambda^\prime )^{2m-2})$ errors. 
To get higher accuracy, higher order terms in the Taylor expansion (9) are needed. 

In the ordinary renormalization procedure, only the terms $\Phi ^n~(n\leq 4)$ should be determined phenomenologically, since the limit $\Lambda^\prime \rightarrow \infty$ is taken. 
However, if $\Lambda$ is finite and $\Lambda^\prime$ can not got to infinity, the errors of the order $(1/\Lambda^\prime )^2$ remain. 
According to Lepage's proposal in QED case [10], we proposed that not only the term $\Phi ^n~(n\leq 4)$ but also the higher terms of $\Phi$ should be determined phenomenologically to remove the errors which arise from the finiteness of $\Lambda^\prime$. 
For an example, we should determine the coefficients for the term $\Phi ^l~(l\leq 6)$, if we want the results with $O((1/\Lambda^\prime )^4)$ error. 
This is equivalent to use 
$$
L^\prime =L-C_5\Phi^5-C_6\Phi^6   \eqno{(11)} 
$$
as the effective Lagrangian with the cutoff $\Lambda^\prime$, instead of the original Lagrangian (1). 
The coefficient $C_5$ and $C_6$ are determined phenomenologically as well as $C_{n}~(n\leq 4)$. 
In that case, the total energy density of the system is given by
$$   \varepsilon 
={{g }\over{(2\pi )^3}}\int^{k_F}_0dk^3\sqrt{{\bf k}^2+M^{*2}}
+{{C_v^2}\over{2M^2 }}\rho^2
+{{M^2}\over{2C_s^2}}\Phi^2
$$
$$+U(\Phi )+U_{1-loop}(M^* ,\Lambda^\prime )+C_5\Phi^5 +C_6\Phi^6.      \eqno{(12)} $$
The error and the $\Lambda^\prime$-dependence of this energy density are the order of $(1/\Lambda^\prime )^4$. 
Similarly, by adding more higher terms of $\Phi^l$ to the right-hand side of eq. (11), we can get the results, the error and the $\Lambda^\prime$-dependence of which are the order of $(1/\Lambda^\prime )^{2j-2}$, where $j=l_{max}/2$. 
( Note that $l_{max}$ should be a even number. ) 

What is the merits of using the new Lagrangian $L^\prime$ with $\Lambda^\prime$ instead of the original Lagrangian $L$ with the original cutoff $\Lambda$? 
The main merits are following. 

\noindent
(1) 
By using the new extended Lagrangian $L^\prime$, the contributions beyond the approximations used in the calculations are also reincorporated phenomenologically into the calculations as well as the contributions which are discarded in using the convenient cutoff $\Lambda^\prime$ instead of $\Lambda$. 

\noindent
(2) 
In the ordinary cutoff theory as is described in section 2 of ref. [7], 
the results may depend on the details of the regulator which is used in the calculations, even if the value of $\Lambda$ are determined phenomenologically. 
However, in the new method, such a dependence on the details of the regulator is removed order by order, by determining the coefficients of the Lagrangian phenomenologically, as well as the dependence on $\Lambda^\prime$ itself. 

\noindent
(3) 
In the Hartree approximations, the momentum dependence of the physical quantities is not calculated. 
However, e.g., in the calculations of the vertex corrections [12], by introducing new cutoff $\Lambda^\prime$, it can be omitted consistently to calculate the momentum dependence of the quantities in high momentum region $(k_E^2 > \Lambda^{\prime 2})$ in which we are not interested. 
This makes the calculations much easier. 

It should be also emphasized that the effective Lagrangian (11) has the "nonrenormalizable" terms, although the original Lagrangian is "renormalizable". 
Of course, this causes no difficulty since we use (11) with the cutoff $\Lambda^\prime$. 
On the other hand, if the "nonrenormalizable" Lagrangian (11) is the original Lagrangian at $\Lambda$, we must determine $C_5$ and $C_6$ phenomenologically as well as $g_s$, $g_v$, $C_0$, $C_1$, $C_2$, $C_3$ and $C_4$. 
This indicates that, in the cutoff field theory, it is difficult to know whether the original Lagrangian is "renormalizable" or "not", unless $\Lambda$ is exactly known. 
In this context, "nonrenormalizable" Lagrangian is natural and useful in the cutoff field theory. 

One may wonder whether the expansion (9) well converges if $\Lambda$ and $\Lambda^\prime$ are not so large as is expected in the nuclear physics. 
In fact, the direct calculation of the expansion (9) shows that the expansion does not well converges if $\Lambda^\prime < 5$GeV [7]. 
However, we can re-expand (9) as
$$  \Delta U=\sum_{l=0}^{\infty} {{1}\over{l!}}{{\partial^l( \Delta U)}\over{\partial \Phi^l}}\vert_{\Phi =0}\Phi^l. \eqno{(13)} $$
Comparing (9) with (13), it is easily seen that $l$-th term in (13) is order of $(1/\Lambda^\prime )^{2j-4}$, where $j$ is the integer part of $(l+1)/2$. 
However, different from the expansion in (9), besides the contribution of the order of $(1/\Lambda^\prime )^{2j-4}$, all higher order contributions which are proportional to $\Phi^{l}$ are also included. 
It has been shown that the expansion (13) well converges even if $\Lambda^\prime$ is not so large [7]. 
It should be remarked that the expansion (13) is automatically used if the coefficients of $\Delta U$ are determined phenomenologically. 
Therefore, we can use the expansion (13) in nuclear physics. 

We also remark that the coefficients such like $C_5$ and $C_6$ are not needed to be observed directly. 
It can be also determined by the physical quantities at the finite density as is in the cases of the other phenomenological models
(See, e.g., [15]). 
For an example, if the effective nucleon mass $M^*$ and the incompressibility $K$ are given in addition to the saturation conditions, 
$C_5$ and $C_6$ can be determined. 
Some examples of the parameter sets are already shown in ref. [7]. 
In section 3, we analyze the compressional properties of the nuclear matter in detail using the effective Lagrangian (11). 

\bigskip

\bigskip


\centerline{\bf 3. Compressional Properties of Nuclear Matter}
\centerline{\bf in Cutoff Field Theory}

\bigskip

In this section, we study the compressional properties of the nuclear matter in detail using the cutoff field theory described in the previous section. 
There are eight parameters in our model. 
If we require the conditions (3) as in the ordinary renormalization procedure, 
four parameters, i.e., $C_s$, $C_v$, $C_5$ and $C_6$ remain as the free parameters of the model. 
If the effective nucleon mass is given at the normal density $\rho_0$, $C_v$ is  determined by the relation [16] 
$$ M^*_0=\sqrt{[M+E_{b0}-C_v^2\rho_0 /M^2 ]^2-k_{F0}^2} \eqno{(14)} $$
which is derived from the Hugenholtz-van Hove theorem [17] at the normal density $\rho_0$. 
The subscript $0$ denotes that the corresponding physical quantity is the one at the normal density $\rho_0$. 
By putting $C_v^2=0$ in eq. (14), it is also seen that the upper limit of the effective nucleon mass $M^{*}_0$ at the normal density is 0.944$M$. 

The other three parameters are determined by giving the saturation condition and the value of the incompressibility 
$$    K=9\rho_0^2{{d^2E_b}\over{d\rho^2}}\vert_{\rho =\rho_0}, \eqno{(15)} $$
where $E_b=\varepsilon /\rho -M$. 

Below, we use $E_{b0}=$-15.75MeV as the binding energy at $\rho_0=0.15$fm$^{-3}$. 
We treat $M^*_0$ and $K$ as variable inputs. 

Using these inputs, we calculate the skewness coefficient which is defined by
$$   K'=3\rho_0^3{{d^3E_b}\over{d\rho ^3}}\vert_{\rho =\rho_0}. \eqno{(16)} $$
In our definition of $K'$, large $K'$ means that the equations of the state becomes hard at high densities. 

In fig. 1, we show the $\Lambda^\prime$ dependence of $K'$. 
In these calculations, first we search the parameter sets of $C_s$, $C_v$, $C_5$ and $C_6$ to reproduce the saturation conditions and the given $M^*_0$ and $K$ at $\Lambda^\prime=$1.5GeV. After that, fixing $C_s$ and $C_v$, we vary $\Lambda^\prime$ and determine $C_5$ and $C_6$ to reproduce the saturation conditions. 
Using new $\Lambda^\prime$ with new $C_5$ and $C_6$, $K^\prime$ is calculated. 
It is expected that the effects of changing $\Lambda^\prime$ are almost canceled by the effects of the change of $C_5$ and $C_6$, if the effective Lagrangian (11) is valid. 

As is seen from these figures, the $\Lambda^\prime$ dependence of $K^\prime$ is very small when $M^*_0=0.75M$ and $0.9M$. 
In the case of $M^*_0=0.6M$, there is the weak $\Lambda^\prime$ dependence. 
This indicates that the higher terms of $\Phi$ in the $\Delta L$, which are neglected, may become important when $\Phi$ is large, i.e., $M^*_0$ is small. 
However, even in these cases above, the dependence is only the order of 10 percents of $K^\prime$ itself. 
The $\Lambda^\prime$ dependence is well removed by the corrections of the order $(1/\Lambda^\prime )^2$ in (11). 
Therefore, we fix $\Lambda^\prime =1.5$GeV below. 

\bigskip


\centerline{$\underline{~~~~~~~}$}

\centerline{Fig. 1 (a), (b) and (c)}

\centerline{$\underline{~~~~~~~}$}


\bigskip

In fig. 2, we show the $K-K'$ relation with fixed $M^*_0$. 
From the figure, it is seen that, in the case of $M^*_0\leq 0.8M$, $K'$ becomes larger as $K$ becomes larger. 
In the case of $M^*_0=0.9M$, $K'$ decreases in the region 180MeV $^<_\sim K$. 
This decrease of $K'$ is related to the rapid decrease of $C_5$. 
The $C_5$ becomes negative in the case of $M^*_0=0.9M$ and $K^>_\sim$ 250MeV, while in the other cases $C_5$ is positive. 

\bigskip


\centerline{$\underline{~~~~~~~}$}

\centerline{Fig. 2}

\centerline{$\underline{~~~~~~~}$}


\bigskip

The $K'$ is related to the coefficients of the leptodermous expansion [18] [19] of nucleus incompressibility $K(A,Z)$ as follows. 
$$  K(A,Z)=K+K_{sf}A^{-1/3}+K_{vs}I^2+K_cZ^2A^{-4/3}+\cdot \cdot \cdot~~~~~;~~~I=1-2Z/A, \eqno{(17)} $$
where the coefficients $K_{sf}$, $K_{vs}$ and $K_c$ are surface term coefficient, volume-symmetry coefficient and Coulomb coefficient, respectively. 
In the scaling model [18][19], $K_c$ is related with $K$ and $K'$ via 
$$ K_c=-{{3e^2}\over{5r_0}}({{9K^\prime}\over{K}}+8),  \eqno{(18)} $$
where $e$ is the electric charge of proton and ${{4\pi}\over{3}}r_0^3=1/\rho_0$. 
Pearson [19] pointed out that there is strong correlation between 
$K$ and $K_c$ which are determined from the experimental information of the giant isoscalar monopole resonances, although, at present, $K$ and $K_c$ are not determined uniquely. 
( He pointed out that $K=120\sim 351$MeV is available to fit the data. )
The relation which Pearson found is shown in table I [19]. 
The $K_{vs}$ which has also strong correlation with $K$ is shown in the table, too. 
The similar results are also gotten by Shlomo and Youngblood [20]. 

\bigskip


\centerline{$\underline{~~~~~~~}$}

\centerline{Table I}

\centerline{$\underline{~~~~~~~}$}


\bigskip

Putting the value of $K'$ into eq. (18), we calculate $K_c$ as the function of $K$. 
The results are shown in fig. 3. 
In the case of $M^*_0\leq 0.8M$, $K_c$ becomes smaller or more negative as $K$ becomes larger. 
In these cases, the calculated values are close to the empirical one at each $K$. 
In the case of $M^*_0=0.9M$, $K_c$ increases in the region $K>220$MeV. 
This increase is related to the decrease of $K'$ in fig. 2, however, the value of $K$ at the starting point of this increase of $K_c$ becomes somewhat larger than the one at the starting point of the decrease of $K'$ by the factor $K$ in the denominator of the right-hand-side of eq. (18). 

\bigskip


\centerline{$\underline{~~~~~~~}$}

\centerline{Fig. 3}

\centerline{$\underline{~~~~~~~}$}


\bigskip

Next we search the parameter sets which reproduce the empirical value of $K$ and $K_c$ in table I. 
If we require $C_6>0$, which ensures the stability of the solution of EOM of $\sigma$-meson, the allowed region of $K$ is restricted. 
The results are summarized in table II(a). 
If this condition are required, only $K=250\pm 50$MeV is allowed. 
In all allowed cases, $M^*_0>0.85M$. 

\bigskip


\centerline{$\underline{~~~~~~~}$}

\centerline{Table II(a) and (b)}

\centerline{$\underline{~~~~~~~}$}


\bigskip

If the negative $C_6$ is not forbidden, some other parameters sets are allowed. 
Such parameter sets which reproduce the mean value of $K_c$ in table I are summarized in table II(b). 
In contrast to the cases with $C_6>0$, $M^*_0<0.85M$ in table II(b). 
Naturally, the solution of EOM of $\sigma$-meson becomes 
unstable at the limit $\Phi \rightarrow \infty$ in the case of negative $C_6$. 
However, since we have neglected the higher terms of $\Phi$ in $\Delta L$, the large $\Phi$ region may be out of scope of this approximation. 
In fig. 4, we show the derivative $\varepsilon^\prime$ of total energy density with respect to $\Phi$. 
The zero point of ${{\partial\varepsilon}\over{\partial\Phi}}$ in the figure is the solution of eq. (7). 
The solution is stable if $\varepsilon^\prime$ is negative for $\Phi < \Phi_s$ and positive for $\Phi > \Phi_s$, where $\Phi_s$ is the solution. 
From the figure, it is seen that the solutions of the parameter sets PS6, PS7 and PS8 are very unstable. 
However, the solutions of the parameter sets PS9 and PS10 are stable in the region $\Phi < M$. 
Therefore, below, we keep the PS9 and PS10 as the candidates of the realistic parameter sets. 
In these cases, $K$ lies in the region 250$\sim$300MeV. 

\bigskip


\centerline{$\underline{~~~~~~~}$}

\centerline{Fig. 4}

\centerline{$\underline{~~~~~~~}$}


\bigskip

In the scaling model, 
volume-symmetry coefficient in eq. (17) is also related to $K$ and $K'$, i.e., 
$$  K_{vs}=K_{sym}-L\biggl( 9{K'\over{K}}+6 \biggr), \eqno{(19)} $$
where
$$  L=3\rho_{0}{{d a_{4}}\over{d \rho}}\vert_{\rho =\rho_{0}},
~~~~~K_{sym}=9\rho_{0}^2{{d^2 a_{4}}\over{d \rho^2}}\vert_{\rho =\rho_{0}},
~~~~~{\rm and}
~~~~~a_{4}={1\over{2}}\rho{{\partial^2\epsilon}\over{\partial\rho_3^2}}\vert_{\rho_3=0}.
  \eqno{(20)}  $$
In eq. (20), $\rho_3=\rho_p-\rho_n$, and $\rho_p$ and $\rho_n$ are proton density and neutron density, respectively. 
It is well known that $\rho$-meson contribution is important in the calculations of these quantities which characterize the symmetric properties of the nuclear matter [21] [13]. 
The $\rho$-meson contribution to the energy density in the Hartree approximation is given by [21] [13] 
$$ \epsilon_\rho ={{g_\rho^2}\over{8m_\rho^2}}(\rho_p-\rho_n)^2={{C_\rho^2}\over{8M^2}}\rho_3^2, \eqno{(21)} $$
where $m_\rho$ and $g_\rho$ are $\rho$-meson mass and $\rho$-nucleon coupling, respectively, and $C_\rho =g_\rho M/m_\rho$. 
We choose $C_\rho$ to reproduce the empirical value of symmetry energy, $a_4=30.0$MeV and calculate $L$, $K_{sym}$ and $K_{vs}$ including the $\rho$-meson contribution (21). 
The results are also summarized in table II and fig. 5. 
The empirical value of sets of $(K,K_{vs})$ [19][20] are also shown in fig. 5. 
From the figure, we see that only parameter set PS4 can well reproduce Pearson data of $K$, $K_c$ and $K_{vs}$ at the same time. 
Among the other parameters set, PS10 gives $K_{vs}$ which is close to the empirical value. 

\bigskip


\centerline{$\underline{~~~~~~~}$}

\centerline{Fig. 5}

\centerline{$\underline{~~~~~~~}$}


\bigskip

In table II(b), for negative $C_6$ we only show the parameter sets which reproduce $K$ and the corresponding mean value of $K_c$ in table I. 
If we use the upper bound of $K_c$, we get other parameter sets. 
Among them, $C_s^2=358.11$, $C_v^2=270.07$, $C_5M=0.0025532$, $C_6M^2=-0.0018579$ and $C_\rho^2=66.135$ which are chosen to reproduce $K=350.0$MeV, $K_c=-7.274+2.06$MeV and $a_4=30.0$MeV, give $M^*_0=0.5442M$ and $K_{vs}=-357.8$. 
The value of $K_{vs}$ has good agreement with the empirical data in table I. 
We call this parameter set PS11. 
This parameter set has the stable solution of EOM of $\sigma$-meson in the region $\Phi< M$. 
On the other hand, the parameter sets which reproduce the lower value of $K_c$ in the table I has a unstable solution of EOM of $\sigma$-meson. 

Finally, we show the $\rho$-dependence of the binding energy and the effective nucleon mass by using the parameter sets PS1$\sim$5, 9, 10 and 11
, which have the stable solution of EOM of $\sigma$-meson in the region $\Phi <M$. 
The results are shown in fig. 6 and fig. 7. 
From fig. 6, it is seen that the equations of state is much more harder at high densities in the cases of PS9, PS10 and PS11 than the other cases with $C_6>0$. 
Samely from fig. 7, we see that the effective nucleon mass becomes much smaller at high densities in the cases of PS9, PS10 and PS11 than the other cases. 
This behavior is related to the small $M^*_0$ in the parameters sets  PS9, PS10 and PS11. 
From eq. (14), small $M^*_0$ means large $C_v$. 
In the parameter sets PS9, PS10 and PS11, the effects of the large $C_v$ overcome the effects of the negative $C_6$, and make the equations of state hard and $M^*$ small at 
high densities. 
It seems that the effects of the $\omega$-nucleon couplings are more important than the effects of $C_6$ at higher densities. 
However, it should be remarked that the higher terms of $\Phi^l$ which can be neglected at the normal density may become important at higher densities, since $\Phi$ becomes larger at that region. 
The effects of the higher terms at the higher densities are open questions.

\bigskip


\centerline{$\underline{~~~~~~~}$}

\centerline{Figs. 6, 7}

\centerline{$\underline{~~~~~~~}$}


\bigskip

\bigskip


\centerline{\bf 4. Summary}

\bigskip

The results obtained in this paper are summarized as follows. 

(1) Using the effective Lagrangian which is constructed in the framework of renormalization group methods, we have studied the compressional properties of the nuclear matter with the cutoff field theory, under the Hartree approximation. 

(2) By including the corrections of the order $(1/\Lambda^\prime)^2$, i.e., $\Phi^5$ and $\Phi^6$ terms, into the effective Lagrangian, 
the physical results become to depend hardly on the value of the convenient cutoff $\Lambda^\prime$ which is conveniently introduced into the theory. 

(3) The compressional properties, the $K$-$K'$, $K$-$K_c$ and $K$-$K_{vs}$ relations in the case of very large $M^*_0(=0.9M)$ have the opposite features to the ones in the case of $M^*_0\leq 0.8M$.

(4) The coefficients of the effective Lagrangian were determined phenomenologically to reproduce the empirical data of $K$ and $K_c$. 

(5) In addition to the requirement of reproducing the empirical data of $K$ and $K_c$, if we require the stability of the solution of EOM of the $\sigma$-meson at the limit $\Phi \rightarrow \infty$, $K=200\sim 300$MeV are favorable. 
However, if the effects of the higher terms of $\Phi$ may become important at large $\Phi$ region, it is out of the scope of this approximation to discuss the stability at $\Phi \rightarrow \infty$. 
If this condition of stability is not absolute in this approximation, the other choices of $K$ are also possible. 

(6) $K=250\sim 350$MeV is favorable to reproduce $K$, $K_c$ and $K_{vs}$ simultaneously. 
Naturally, this result is consistent with the result of the general discussion which does not depend on the details of the $\sigma$-meson potential [22]. 

(7) It seems that the effect of the $\omega$-nucleon coupling is more important than the effect of $C_6$ at higher densities. 
However, the higher terms $\Phi^l$ which can be neglected at the normal density may become important at higher densities, since $\Phi$ becomes larger at that region. 
The effects of the higher terms at the higher densities are open questions. 

It is very interesting to reexamine the calculations of the meson self-energy in the random phase approximation by including the $(1/\Lambda^\prime )^2$ corrections with the parameter sets obtained by this paper. 
It is now under the studies. 

\noindent
Acknowledgement: The authors gratefully thank K. Harada, who recommended them to use the renormalization group methods in quantum hadrodynamics, for many useful suggestions and discussions. 
They also thank T. Kohmura, T. Suzuki, M. Yahiro and H. Yoneyama for useful discussions, 
 and acknowledge the computing time granted by Research Center for Nuclear Physics (RCNP). 

\vfill\eject

\centerline{\bf References}


\noindent
[1] J.D. Walecka, Ann. of Phys. {\bf 83} (1974) 491

\noindent
[2] S.A. Chin, Phys. Lett. {\bf 62B} (1976) 263: S.A. Chin, Ann. of Phys. {\bf 108} (1977) 301

\noindent
[3] T. Kohmura, Y. Miyama, T.Nagai, S. Ohnaka, J. Da Provid{\^{e}}ncia, and T. Kodama, Phys. Lett. {\bf B226}(1989)207. 

\noindent
[4] R.J. Furnstahl and C.J. Horowitz, Nucl. Phys. {\bf A485} (1988) 632.

\noindent
[5] H. Kurasawa and T. Suzuki, Nucl. Phys. {\bf A445} (1985) 685: 

\noindent
H. Kurasawa and T. Suzuki, Nucl. Phys. {\bf A490} (1988) 571. 

\noindent
[6] T.D. Cohen, Phys. Lett. {\bf B211}(1988)384. 

\noindent
[7] H. Kouno, T. Mitsumori, Y. Iwasaki, K. Sakamoto, N. Noda, K. Koide, A. Hasegawa and M. Nakano, Prog. Theor. Phys., {\bf 97} (1997)91. 

\noindent
[8] K.G. Wilson and J. Kogut, Phys. Rep. {\bf 12} (1974)75. 

\noindent
[9] K. G. Wilson, Rev. Mod. Phys. {\bf 55} (1983)583. 

\noindent
[10] G P. Lepage, "What is renormalization?" in "From actions to answers", Proceedings of the 1989 theoretical advanced study institute in elementary particle physics, edited by T. DeGrand and D. Toussaint, p483, (World Scientific, Singapore 1990). 

\noindent
[11] M. Nakano, N. Noda, T. Mitsumori, K. Koide, H. Kouno and A. Hasegawa, 
to be published in Physical Review C. 

\noindent
[12] H. Kouno, K. Koide, N. Noda, K. Sakamoto, Y. Iwasaki, T. Mitsumori, A. Hasegawa and M. Nakano, preprint SAGA-HE-112-97. 

\noindent
[13] 
B.D. Serot and J.D. Walecka, $The$ $Relativistic$ $Nuclear$ $Many$-$Body$ $Problem$ in: Advances in nuclear physics, vol. {\bf 16}, edited by J.W. Negele and E. Vogt (Plenum Press, New York, 1986).

\noindent
[14] E.K. Heide and S. Rudaz, Phys. Lett. {\bf B262}(1991)375. 

\noindent
[15] H. Kouno, N. Kakuta, N. Noda, K. Koide, T. Mitsumori, A. Hasegawa and M. Nakano, Phys. Rev. {\bf C51} (1995) 1754: 

\noindent
H. Kouno, K. Koide, T. Mitsumori, N. Noda, A. Hasegawa and M. Nakano, Phys. Rev. {\bf C52} (1995) 135: 

\noindent
H. Kouno, K. Koide, T. Mitsumori, N. Noda, A. Hasegawa and M. Nakano, Prog. Theor. Phys. {\bf 96} (1996) 191. 

\noindent
[16] J. Boguta and and A.R. Bodmer, Nucl. Phys. {\bf A292}(1977)413. 

\noindent
[17] N.M. Hugenholtz and L.van Hove, Physica {\bf 24} (1958)363.

\noindent
[18] J.P. Blaizot, Phys. Rep. {\bf 64} (1980) 171. 

\noindent
[19] J.M. Pearson, Phys. Lett. {\bf B271} (1991) 12. 

\noindent
[20] S. Shlomo and D.H. Youngblood, Phys. Rev. {\bf{C47}} (1993)529.

\noindent
[21] B.D. Serot, Phys. Lett. B{\bf{86}} (1979)146

\noindent
[22] H. Kouno, T. Mitsumori, N. Noda, K. Koide, A. Hasegawa and M. Nakano, Physical Review {\bf C53} (1996) 2542. 


\noindent

\vfill\eject

\centerline{\bf Table and Figure Captions}

\bigskip

\bigskip

\noindent
Table I The sets of the empirical values of $K$, $K_c$ and $K_{vs}$ from the Table 3 in ref. [19]. 
According to the conclusion in ref. [19],  only the data in the cases of $K=150\sim 350$MeV are shown. )

\bigskip

\bigskip

\noindent
Table II The parameter sets which reproduce the empirical value of $K$ and $K_c$ in table I. 
$K$, $K_c$, $K'$, $L$, $K_{sym}$ and $K_{vs}$ are shown in MeV. 
(a) The sets with positive $C_6$ which reproduce the parameter sets of the values of $K$ and the corresponding upper limits, mean values and lower limits of $K_c$ in table I. 
(b) The sets with negative $C_6$ which reproduce the values of $K$ and the corresponding mean values of $K_c$ in table I. 

\bigskip

\bigskip

\noindent
Fig. 1 The $\Lambda^\prime$-dependence of $K'$. 
The solid, dotted and dashed lines correspond to the case of $K=$150, 250 and 350MeV, respectively. 
(a) The results with $M^*_0=0.9M$. 
(b) The results with $M^*_0=0.75M$. 
(c) The results with $M^*_0=0.6M$. 

\bigskip

\bigskip

\noindent
Fig. 2 $K$-$K'$ relation in the case with $\Lambda^\prime=$1.5GeV. 
The solid, dotted, dashed, dashed-dotted and bald solid lines correspond to the results with $M^*_0 =0.5M$, $0.6M$, $0.7M$, $0.8M$ and $0.9M$, respectively. 

\bigskip

\bigskip

\noindent
Fig. 3 $K$-$K_c$ relation in the case with $\Lambda^\prime=$1.5GeV. 
Each line has the same correspondence to $M^*_0$ as in fig. 2. 
The crosses with error bars are the empirical data in table I, while the small solid triangles are the empirical data from Table IV in ref. [20]

\bigskip

\bigskip

\noindent
Fig. 4 $\varepsilon^\prime =\partial \varepsilon/\partial \Phi$ is shown as the function of $\Phi$. 
The zero-point is the solutions of EOM of the $\sigma$-meson. 
The solid, dotted, dashed, dashed-dotted and bald solid lines correspond to the results with the parameter sets, PS6, PS7, PS8, PS9 and PS10, respectively. 

\bigskip

\bigskip

\noindent
Fig. 5 $K$-$K_{vs}$ relation in the case with $\Lambda^\prime=$1.5GeV. 
Each line has the same correspondence to $M^*_0$ as in fig. 2. 
The open circle at $K=250$MeV, the open inverse triangles at $K=250$ and 300MeV, the open triangles at $K=200$ and 250MeV, the solid circles at $K=150$, 200, 250 and 300MeV, the solid square at $K=250$MeV, and the solid inverse triangle at $K=350$MeV are the results with the parameter sets PS3, PS2, PS5, PS1, PS4, PS6, PS7, PS8, PS10, PS9 and PS11, respectively. 
The crosses with error bars and the small solid triangles have the same correspondence to the empirical data as in fig. 3. 

\bigskip

\bigskip

\noindent
Fig. 6 The binding energy $E_b$ is shown as the function of the baryon number density $\rho$. 
The solid, dotted, dashed, dashed-dotted, bald solid, bald dotted, bald dashed lines  and the small solid circles correspond to the results with the parameter sets, PS1, PS2, PS3, PS4, PS5, PS9, PS10 and PS11, respectively. 

\bigskip

\bigskip

\noindent
Fig. 7 The effective mass of nucleon $M^*$ is shown as the function of the baryon number density $\rho$. 
Each line and the small solid circles have the same correspondence to the parameter sets in fig. 6. 

\bigskip

\bigskip

\vfill\eject

\large

\hspace*{-4cm}
  \begin{tabular}{cccccc}
                                    \hline 
    \     & Set 1 & Set 2 & Set 3 & Set 4 & Set 5  \\ \hline
    \   $K$ & 150.0 & 200.0  & 250.0 & 300.0 & 350.0 \\
    \  $K_c$ & $5.861\pm2.06 $ & $2.577\pm2.06 $ & $-0.7065\pm2.06 $
     & $-3.990\pm2.06 $ &$-7.274\pm2.06 $ \\
    \ $K_{vs}$ & $66.83\pm101$ & $-46.94\pm101$ & $-160.7\pm101$
     & $-274.5\pm101$ & $-388.3\pm101$ \\ \hline
  \end{tabular}

\bigskip

\begin{center}
Table I
\end{center}


\bigskip

\bigskip

\bigskip

\vfill\eject

\large

\hspace*{-3cm}
  \begin{tabular}{cccccc}
                                    \hline 
    \     & PS1 & PS2 & PS3 & PS4 & PS5  \\ \hline
    \   $K$ & 200.0 & 250.0  & 250.0 & 250.0 & 300.0 \\
    \  $K_c$ & 2.577-2.06 & -0.7065+2.06 & -0.7065
     & -0.7065-2.06 & -3.990+2.06 \\
    \  $K'$ & -193.3 & -273.1 & -195.7
     & -118.2 & -179.6 \\
    \  $C_s^2$ & 52.840 & 11.146 & 83.165
     & 124.96 & 106.33 \\
    \  $C_v^2$ & 5.9060 & 4.5616 & 17.407
     & 38.537 & 32.331 \\
    \  $C_5M$ & -72.950 & -668.67 & -9.9886
     & 0.52416 & -1.3005 \\
    \  $C_6M^2$ & 968.51 & 8199.5 & 121.44
     & 0.53427 & 16.321 \\
    \  $C_{\rho}^2$ & 110.56 & 110.70 & 109.33
     & 106.97 & 107.68 \\
    \  $M^*_0/M$ & 0.9358 & 0.9378 & 0.9192
     & 0.8887 & 0.8976 \\
    \  $L$ & 77.40 & 77.14 & 77.57
     & 78.36 & 77.88 \\
    \  $K_{sym}$ & -26.55 & -26.40 & -27.49
     & -26.91 & -28.60 \\
    \ $K_{vs}$ & 182.4 & 269.1 & 53.48
     & -163.5 & -76.22 \\ \hline
  \end{tabular}

\bigskip

\begin{center}
Table II(a)
\end{center}


\bigskip

\bigskip

\bigskip

\vfill\eject

\large

\hspace*{-3cm}
  \begin{tabular}{cccccc}
                                    \hline 
    \     & PS6 & PS7 & PS8 & PS9 & PS10  \\ \hline
    \   $K$ & 150.0 & 200.0  & 250.0 & 250.0 & 300.0 \\
    \  $K_c$ & 5.861 & 2.577 & -0.7065
     & -0.7065 & -3.990 \\
    \  $K'$ & -265.5 & -255.3 & -195.7
     & -195.7 & -86.69 \\
    \  $C_s^2$ & 171.46 & 196.62 & 205.79
     & 382.13 & 378.50 \\
    \  $C_v^2$ & 78.359 & 107.53 & 118.82
     & 290.93 & 288.38 \\
    \  $C_5M$ & 0.24751 & 0.072648 & 0.043383
     & 0.0025779 & 0.0024938 \\
    \  $C_6M^2$ & -0.78428 & -0.17793 & -0.092024
     & -0.0019553 & -0.0018217 \\
    \  $C_{\rho}^2$ & 102.08 & 98.082 & 96.431
     & 60.237 & 60.991 \\
    \  $M^*_0/M$ & 0.8308 & 0.7881 & 0.7715
     & 0.5116 & 0.5156 \\
    \  $L$ & 80.77 & 82.28 & 82.76
     & 108.8 & 107.5 \\
    \  $K_{sym}$ & -1.857 & 5.897 & 3.671
     & 260.2 & 227.1 \\
    \ $K_{vs}$ & 800.3 & 457.4 & 90.07
     & 373.8 & -138.3 \\ \hline
  \end{tabular}

\bigskip

\begin{center}
Table II(b)
\end{center}

\end{document}